\documentclass[final,journal,letterpaper]{IEEEtran}

 \usepackage{varioref} 
 \usepackage{wrapfig} 
 \usepackage{threeparttable} 
 \usepackage{dcolumn} 
  \newcolumntype{d}{D{.}{.}{-1}}
 \usepackage{nomencl} 
  \makeglossary
 \usepackage{subfigmat} 
 \usepackage{fancyvrb} 
  \fvset{fontsize=\footnotesize,xleftmargin=2em}
 \usepackage{lettrine} 
 \usepackage[hidelinks]{hyperref}

 \usepackage{epsfig}
 \usepackage{epstopdf}
 \usepackage{subfigure}
 \usepackage{latexsym}
 \usepackage{amsmath}
 \usepackage{amsfonts}
 \usepackage{amssymb}
 \usepackage{mathrsfs}
 \usepackage{cases}
 \usepackage{array}
 \usepackage{color}
  \usepackage{flushend}

 \usepackage{url}
 \usepackage[noadjust]{cite}
 \usepackage{graphicx}
 \usepackage{psfrag}
 \usepackage{color}

\newfont{\Bb}{msbm10 scaled\magstep1}

\begin{document}

\title{Convolutional Neural Network and Transfer Learning for High Impedance Fault Detection}

\author{Rui Fan, Tianzhixi Yin 
\thanks{R. Fan, and T. Yin are with Pacific Northwest National Laboratory, Richland, WA 99354, USA (e-mail: \{Rui.Fan, Tianzhixi.Yin\}@pnnl.gov).}}

\maketitle

\begin{abstract}
This letter presents a novel high impedance fault (HIF) detection approach using a convolutional neural network (CNN). Compared to traditional artificial neural networks, a CNN offers translation invariance and  it can accurately detect HIFs in spite of  variance and noise in the input data.
A transfer learning method is used to address the common challenge of a system with little training data.
Extensive studies have demonstrated the   accuracy and effectiveness of using a CNN-based approach for HIF detection.
\end{abstract}

\begin{IEEEkeywords}
Convolutional neural network, transfer learning, high impedance fault, deep learning
\end{IEEEkeywords}

\section{Introduction}\label{sec:Intro}
\IEEEPARstart{H}{igh} impedance fault (HIF) is detrimental to public safety, and the arcing increases a fire hazard. Protecting distribution systems from HIF is very challenging, as the fault current is usually too low to be detected by over-current relays \cite{review}. Among the advanced methods that have been proposed to resolve this challenge, the artificial neural network (ANN) is well known for its high accuracy in pattern classification and generalization \cite{ANN2}. Existing ANN-based schemes often use multilayer perceptron (MLP) neural networks to detect HIFs in distribution systems \cite{ANN1}. However, traditional ANN-based schemes have two major limitations. First, a traditional MLP does not take into account the spatial structure of data, and it is less effective in obeying translation invariance; thus, variance (noise, length, pattern position) in the input data would affect   its performance in detecting HIFs \cite{Sabisch98}. Second, training an ANN system requires copious data, it is very difficult to use ANN-based schemes in a distribution system with little training data.

To overcome these challenges, we plan to use a convolutional neural network (CNN)  to replace the traditional MLP, and use  the transfer learning technique to address the lack-of-data issue \cite{CNN, Transfer}.
The convolutional layers and pooling layers in a CNN model help preserve the translation invariance; thus, it can accurately detect an HIF even it varies in some way. Once a CNN is well trained, it can effectively identify the edges and patterns of HIFs. Then we can use this  CNN model in a new   system  through transfer learning. Although the   waveforms are not the same in different distribution systems, certain characteristics and patterns (such as the sudden charge or shape) of the HIF  remain unchanged. The previous CNN model has already learned these characteristics and patterns well; thus, fewer data are needed to adapt to the new system. In this letter, we demonstrate the effectiveness and accuracy of using CNN and transfer learning for HIF detection through extensive studies.
 

\section{Proposed Approach}\label{sec:Approach}
To overcome the two aforementioned practical challenges of using traditional ANN-based schemes for HIF detection, we first train a CNN model using extensive simulated training data. The performance of the CNN-based scheme is compared with an MLP-based scheme using the same training and testing data. Then the CNN model is applied to a new distribution system with far fewer data through transfer learning.
\subsection{High Impedance Fault Model}
\begin{figure}[!ht]
\centering
\includegraphics[width=0.31\textwidth]{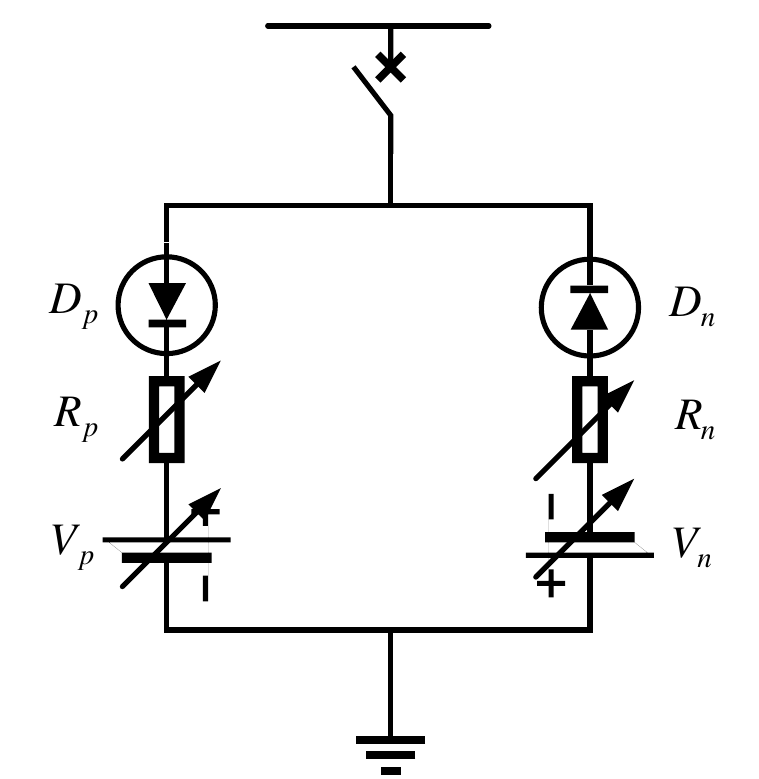}
\caption{HIF model proposed in \cite{HIF1, HIF2}}
\label{fig:1}
\end{figure}

An HIF occurs when an energized   conductor  contacts the ground or a quasi-insulating object, such as a vegetation, buildings or equipment. 
HIFs have some characteristics (such as arc, waveform asymmetry, and non-linearity) that can help CNN-based schemes   differentiate them from load changes or other transients caused by normal operations. In this letter, we use a classical HIF model \cite{HIF1, HIF2} for numerical studies, as shown in Fig. \ref{fig:1}. 

Two anti-parallel connected DC sources  are separately in series with a   diode   and variable resistor, where the subscripts $p, n$ stand for positive and negative properties.
 Thus ,when the instantaneous phase voltage  is larger than the positive voltage $V_p$, the positive cycle of fault current flows towards the ground through the left-hand path in Fig. \ref{fig:1}. Oppositely, when the   voltage  is smaller than the negative voltage $V_n$, the negative cycle of fault current flows through the right-hand path.

\subsection{Convolutional Neural Network}
\begin{figure}[!ht]
\centering
\includegraphics[width=0.42\textwidth]{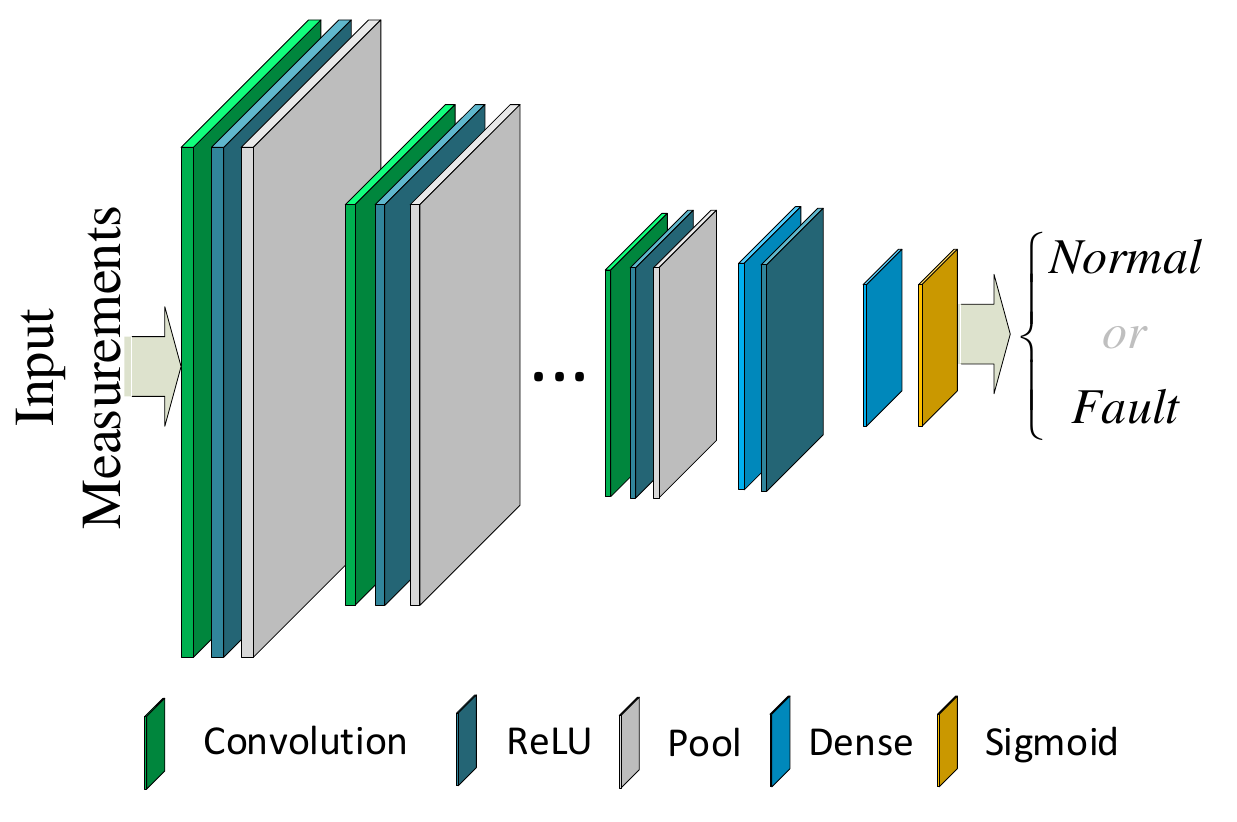}
\caption{CNN-based scheme for HIF detection}
\label{fig:2}
\end{figure}

In recent years, the CNN has achieved remarkable success in various research fields \cite{CNN} because it has many advantages over traditional machine learning approaches such as MLP. A CNN can understand patterns at different levels, which means it learns the details and the overview at the same time. 
A 1D CNN model for HIF detection is shown in Fig. \ref{fig:2}. The input measurements are those of    current waveforms from the bus feeder. The CNN model has four layers, and each layer consists of convolution, rectified linear unit (ReLU), and max-pooling functions. The convolution function helps capture detail patterns, and in a CNN, it is

\begin{equation}
\left( f * g \right) [n] = \sum_i^n{f[i] \times g[i]}
\end{equation}
where $f$ is the filter feature, $g$ is the input that corresponds to the filter, and $n$ is the size of the filter. The output of convolution operation $x$ is passed to a ReLU activation, which it is defined as

\begin{equation}
r(x) = x^+ = max (0,x)
\end{equation}

ReLU handles the vanishing-gradient problem quite well, and it is less computationally expensive than traditional tanh and sigmoid activation operations. A max-pooling operation follows the ReLU activation to reduce the dimensionality and allow    filters in   deeper layers to learn a general overview of the input patterns.  At the end layer of the CNN, a fully-connected layer and an output layer using a sigmoid activation function is used to determine whether the input measurements are HIFs or normal transients. The sigmoid function is 

\begin{equation}
S(x)=\dfrac{e^x}{e^x + 1}
\end{equation}

After obtaining the output $\hat{y}$, the cross-entropy is used as the loss function, which is defined as

\begin{equation}
J = - \dfrac{1}{m} \sum_j^m \left[ y^{(j)} log(\hat{y}^{(j)}) + (1-y^{(j)}) log(\hat{y}^{(j)}) \right] 
\end{equation}
where $y$ is the true label of the input, $m$ is the number of input data, and the superscript $(j)$ stands for the $j^{th}$ observation. The CNN updates its parameters (or weights) of each layer with the objective of decreasing the loss $J$ through a back-propagation process \cite{Backprop}. 

\subsection{Transfer Learning Method}
Training a neural network usually requires large amounts of data. However, in many cases, the number of data is limited; thus, this problem is a common barrier for ANN-based schemes in practical applications.
To address this problem, people have invented transfer learning \cite{Transfer}, which is one of the most powerful concepts in the deep learning realm. 
The basic concept of transfer learning is that instead of training a neural network from the beginning (random initialization), we can leverage the   learning experience  from previous tasks and transfer it to the  training process in a new system.

In this letter, we first trained the CNN model in a system with abundant   data; thus, it was able to capture the characteristics and patterns  of HIFs.  Then we   used this  CNN model with the same  parameters in a new distributed system  through transfer learning. Because transfer learning was like fine-tuning the  CNN model with minimal changes, it   required less data to adapt to the new system, and the training process was usually much faster (than starting from scratches). 

\section{Test Results}\label{sec:Results}

In the first   case, we compared the performance of a CNN-based scheme with an MLP-based scheme in detecting HIFs using the IEEE 34-Bus feeder system. We  generated thousands of HIF data (with different random variables $V_p, V_n, R_p, R_n$), as well as thousands of normal transients (load change, switching, capacitor bank change, etc.). We also used variable loading levels, phases,  locations, and fault/transient inception angles. The HIF fault resistor varied between 100 and 600 Ohms. In addition, 2\% white noise was added to the data to mimic   real-world situations. The sampling rate was 15 kHz, and the input data contained around 300 samples. Both the CNN and the MLP had four layers and used a sigmoid activation function to predict the final output. We randomly selected around 5,000 data (half HIFs and half normal transients), of which 80\% were used for training/validation and 20\% were used for testing.

\begin{figure}[!ht]
\centering
\includegraphics[width=0.48\textwidth]{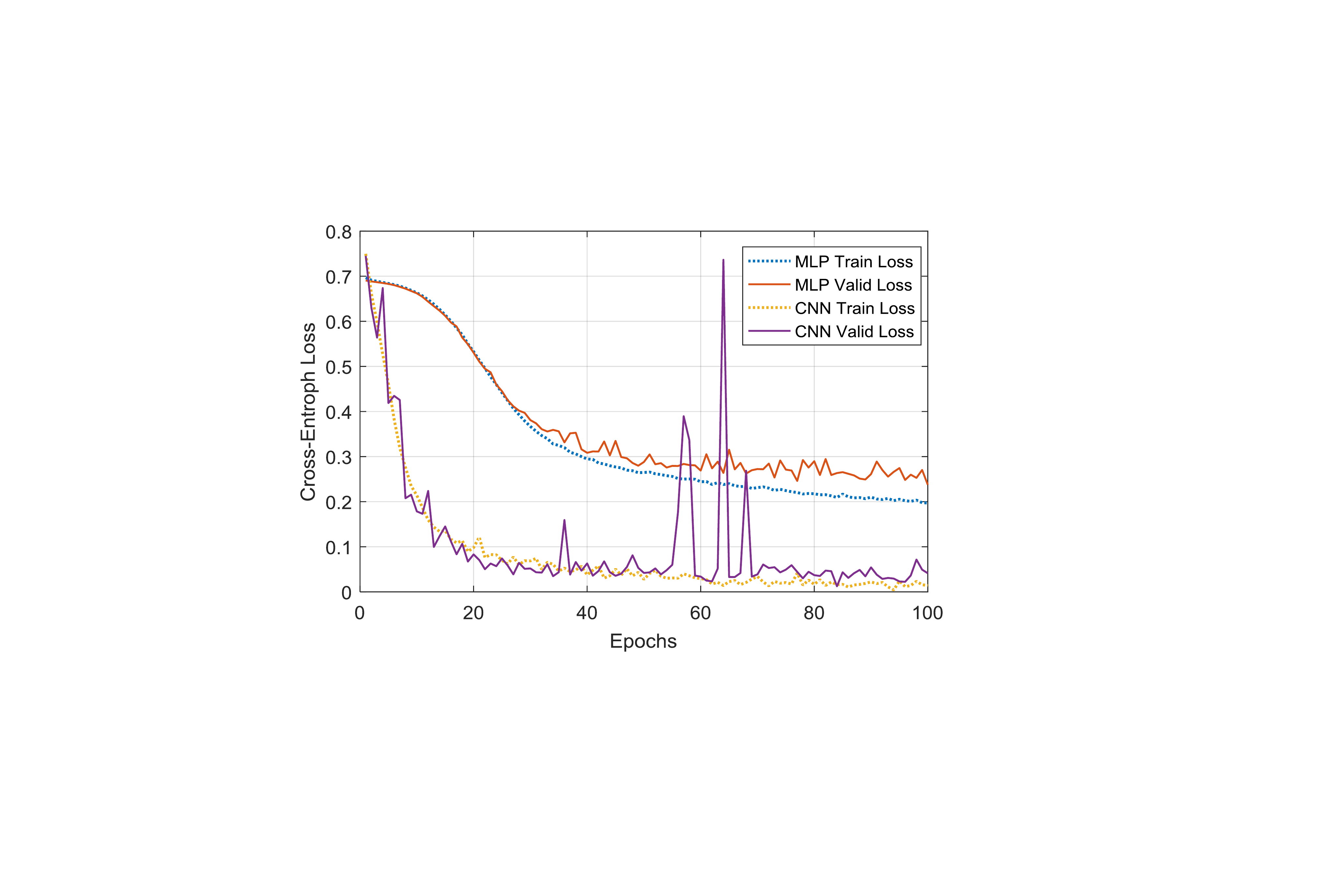}
\caption{Cross-entropy loss of MLP-based and CNN-based scheme}
\label{fig:3}
\end{figure}

The cross-entropy losses of MLP-based and CNN-based schemes during training and validation   are shown in Fig. \ref{fig:3}. It has shown that the cross-entropy losses of both schemes decreased as the training continued, until they finally reached   steady states. It is clear that the loss with an MLP-based scheme is greater than that with a CNN-based scheme, indicating that the MLP-based scheme cannot compete with the CNN-based scheme in differentiating HIFs from normal transients. The test results shown in Table. \ref{table:1} validate  this assumption: the MLP-based scheme has an accuracy of 91.13\% and the CNN-based scheme has a much higher accuracy of 99.52\%.

\emph{Remark 1}: Over-fitting had not occurred in either the MLP-based or the CNN-based scheme, since the training and validation losses matched each other. Also,   a few sparks occurred during the CNN training process, especially when the CNN was close to the minimum loss, because at those moments, the CNN had already     closely approached the optimal points.   It thus tried a direction, discovered that was the wrong way, and   returned to the previous region.

\begin{table}[!ht]
\renewcommand{\arraystretch}{1.2}
\caption{Test results for CNN and MLP in the IEEE 34-Bus feeder}
\label{table:1}
\centering
\begin{tabular}{c|c|c}
\hline
& \textbf{MLP-based} & \textbf{CNN-based}\\
\hline
True Positive & 489/526 & 523/526 \\
\hline
False Positive & 37/526 & 3/526 \\
\hline
False Negative & 55/511 & 2/511 \\
\hline
True Negative & 456/511 & 509/511 \\
\hline
\hline
\textbf{Accuracy} & 91.13 \% & 99.52 \% \\
\hline
\end{tabular}
\end{table}

In the second  case, we tested the transfer learning method in a new distribution system, the IEEE 13-Bus feeder. This time we had much less data ($\approx$300 measurements) than in the previous case. We then compared the same CNN model with these data (50\% for training and validation, 50\% for testing), with the only difference being whether the CNN model was trained from scratches (random initialization) or it leveraged the previous knowledge through transfer learning.

\begin{figure}[!ht]
\centering
\includegraphics[width=0.48\textwidth]{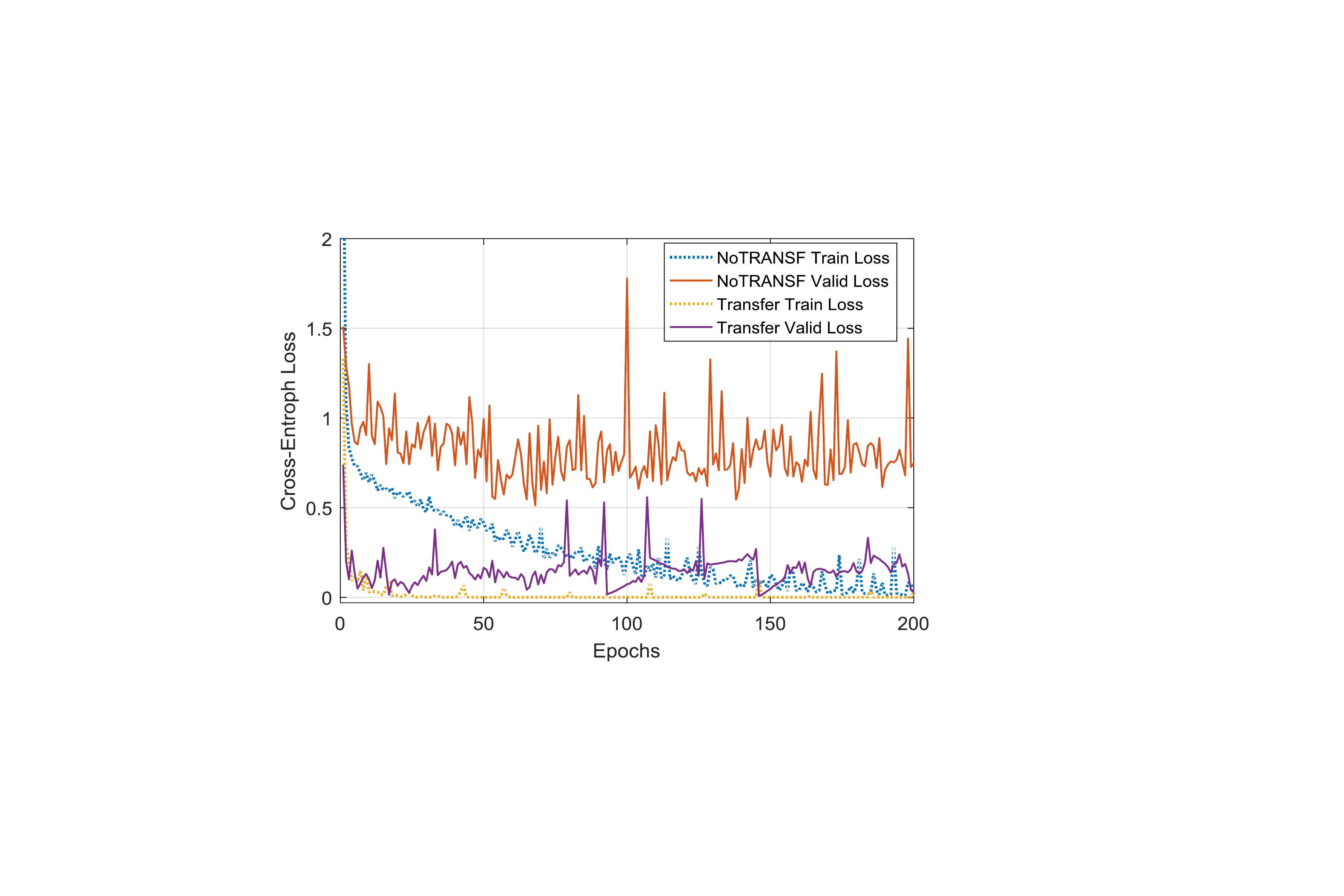}
\caption{Cross-entropy loss of transfer learning and random initialization}
\label{fig:4}
\end{figure}

\begin{table}[!ht]
\renewcommand{\arraystretch}{1.2}
\caption{Test results for CNN with random initialization and transfer learning in the IEEE 13-Bus feeder}
\label{table:2}
\centering
\begin{tabular}{c|c|c}
\hline
& \textbf{Random initialization} & \textbf{Transfer learning}\\
\hline
True Positive & 45/68 & 62/68 \\
\hline
False Positive & 23/68 & 6/68 \\
\hline
False Negative & 18/94 & 2/94 \\
\hline
True Negative & 76/94 & 92/94 \\
\hline
\hline
\textbf{Accuracy} & 74.69 \% & 95.06 \% \\
\hline
\end{tabular}
\end{table}

The cross-entropy losses   during training and validation processes are shown in Fig. \ref{fig:4}. Over-fitting has occurred in the CNN with random initialization, since the loss in training is much less than the loss in validation. In contrast, CNN with transfer learning results in lower losses in both training and validation. Test results for the CNN with random initialization and transfer learning are shown in Table. \ref{table:2}. The CNN with transfer learning provides a much higher accuracy (95.06\%) than   random initialization (74.69\%).

\emph{Remark 2}: Note that the accuracy in Case 2 is lower than that in Case 1; that is because we have far fewer data for   training the CNN model. In addition to the high accuracy achieved by using transfer learning, the CNN model reached   steady state much faster ( in less than 20 epochs). In contrast, training a CNN from scratches took hundreds of epochs to reach a steady state. Therefore, transfer learning can help reduce the computational burden significantly.

\section{Conclusion}\label{sec:Conc}
This letter proposes a CNN and transfer learning based approach to detecting HIFs in distribution systems. Extensive studies have demonstrated the accuracy of the CNN-based scheme in differentiating HIFs from normal transients. Also, the well-trained CNN model can be easily adapted to new systems with little data through transfer learning, which results in higher   accuracy and less computational cost.

\bibliographystyle{IEEEtran}
\bibliography{CNN}

\vfill

\end{document}